\documentclass{aa}
\usepackage{graphics}
\usepackage{psfig}

\newcommand{\oii}{O{\sc ii}}

\newcommand{\nii}{N{\sc ii}}

\newcommand{\sii}{Si{\sc ii}}
\newcommand{\hi}{H\,{\sc i}}
\newcommand{\hii}{H\,{\sc ii}}

\newcommand{\etal}{et al.}

\newcommand{\Ein}{{\em Einstein}}

\newcommand{\ros}{{\em ROSAT}}
\newcommand{\Ros}{{\em ROSAT}}
\newcommand{\ltsim}{\raisebox{-1mm}{$\stackrel{<}{\sim}$}}
\newcommand{\gtsim}{\raisebox{-1mm}{$\stackrel{>}{\sim}$}}

\newcommand{\dgr}{$^{\circ}$}

\newcommand{\eg}{e.g. }
\newcommand{\ie}{i.e. }

\def\arcm{\hbox{$^\prime$}}

\begin{document}

   \thesaurus{11;(11.03.4 Sculptor;
		11.09.1 NGC~7793; 
		11.09.4; 
		11.19.2; 
		13.25.2)}

   \title{ROSAT observations of the Sculptor galaxy NGC~7793}
   \author{A.M.~Read \and W.~Pietsch}
   \offprints{A.M.~Read, e-mail address aread@mpe.mpg.de}
   \institute{Max-Planck-Institut f\"ur extraterrestrische Physik,
              Gie\ss enbachstra\ss e, D--85748 Garching, Germany}
   \date{Received date; accepted date}
   \maketitle
   \markboth{A.M.~Read \& W.~Pietsch: ROSAT observations of the Sculptor galaxy NGC~7793}{}

\begin{abstract}

We report here the results of spectral imaging observations with the \Ros\ PSPC of
the field surrounding the very nearby ($D=3.38$\,Mpc) Sculptor galaxy, NGC~7793.
Many point sources are detected within the field, several of them showing evidence
for variability. Seven sources are seen to lie within the optical confines of the
system, the brightest ($L_{X} \approx 9\times10^{38}$\,erg s$^{-1}$), lying to the
southern edge of NGC~7793. This source, also observed to be variable, is seen now,
not to be associated with a $z=0.071$ redshift QSO, as was previously thought. A 
number of the other sources within the NGC~7793 disc are likely to be due to 
X-ray binaries, supernova remnants or superbubbles within the galaxy itself. Other
sources may be associated with background AGN. In addition to the point source
emission, rather uniform unresolved emission is detected in and around NGC~7793
extending to a radius of perhaps 4\,kpc. This emission is likely to be contaminated
to some degree by unresolved point sources, as its temperature ($kT\approx1$\,keV)
is rather hotter than is seen for the diffuse gas components in other nearby spiral
galaxies. Comparing the X-ray properties of NGC~7793 with those of the remaining 
Sculptor group members, suggests that it may, in terms of activity, lie 
somewhere between its quiescent and starburst neighbours. 


\end{abstract}

\section{Introduction}
\label{sec_intro}

NGC~7793 is a reasonably face-on galaxy ($i =$53.7\dgr), of type Sd(s)
(\rm{IV}) (Sandage \& Tammann \cite{Sandage}), and is one of the five
well-known galaxies making up the nearby Sculptor galaxy group. It has a very
thin bulge and a filamentary spiral structure, though this spiral pattern is
almost lost within the numerous regions of \hii\ emission and star formation.
Though estimates of its distance vary, ranging from 2.5 (de Vaucouleurs \etal\
\cite{RC3}) to 3.4\,Mpc (Carignan \cite{Carignan85}), recent efforts have
concentrated towards the high end of this range, and we here adopt a distance
of 3.38\,Mpc, in agreement with several other authors (Blair \& Long
\cite{Blair}; Carignan \& Puche \cite{Carignan90}; Puche \& Carignan
\cite{Puche91}). At this distance, 1\arcm\ corresponds to just under a
kiloparsec (0.98\,kpc). It is an intrinsically small and dim system,
with an exponential scale length $\alpha^{-1} \simeq 1$, and an absolute $B$
magnitude $M_{B}^{0,i} \simeq -18.3$. This absolute magnitude corresponds to a
total blue luminosity of $3.1\times10^{9} L_{B\odot}$ (Carignan \& Puche
\cite{Carignan90}).

As far as previous X-ray observations of NGC~7793 go, 
NGC~7793 was observed with the \Ein\ IPC for almost 2\,ks in 1979 (Fabbiano
\etal\ \cite{Fabbiano92}). Very little was gleaned from these observations apart
from the fact that the `X-ray centroid does not coincide with the nucleus'.
These observations are described more fully throughout this paper. The \Ros\
X-ray telescope (XRT) however, with the Position Sensitive Proportional Counter
(PSPC) (Pfeffermann \etal\ \cite{Pfeffermann}) at its focal plane, offers three
very important improvements over previous X-ray imaging instruments (such as
the \Ein\ IPC). Firstly, the spatial resolution is very much improved, the 90\%
enclosed energy radius at 1\,keV being $27''$ (Hasinger \etal\
\cite{Hasinger}). Secondly, the PSPC's spectral resolution is very much better
($\Delta E/E \sim 0.4$ FWHM at 1\,keV) than earlier X-ray imaging instruments,
allowing the derivation of characteristic source and diffuse emission
temperatures. Lastly, the PSPC internal background is very low
($\sim3\times10^{-5}$\,ct s$^{-1}$ arcmin$^{-2}$; Snowden \etal\
\cite{Snowden94}), thus allowing the mapping of low surface brightness emission 
source populations 
(see Tr\"{u}mper (\cite{Trumper}) for a description of the
\Ros\ satellite and instruments).

Here we report the results of a 23.8\,ks \Ros\ PSPC observation of the field
surrounding NGC~7793. The plan of the paper is as follows. Sect.\,\ref{sec_obs_data}
describes the observation and the preliminary data reduction methods used,
Sect.\,\ref{sec_disc_sources} discusses the results as regards the point source
emission, and Sect.\,\ref{sec_disc_unres} discusses the results as regards the remaining
unresolved emission. Sect.\,\ref{sec_disc_n7793} discusses the X-ray properties of
NGC~7793, with regard both to its membership of the Sculptor group and to how it
compares to spiral galaxies in general. Finally a summary is presented in
Sect.\,\ref{sec_summary}.

\section{ROSAT observations and preliminary analysis}
\label{sec_obs_data}

The field surrounding NGC~7793 was observed with the \Ros\ PSPC in essentially two
separate observations (on December 7th 1992, and between May 20th-27th 1993) for a
total of 23.8\,ks. Towards the end of the December 1992 observation, a modest increase
(by a factor of $\sim$1.5) in the count rate is observed. Investigating further,
one sees that this extra emission pervades the entire field of view, and is merely an
enhancement in the background over the whole detector. This background enhancement,
which merely decreases the signal to noise ratio over parts of the observation, is noted
here, and its effects are discussed throughout the paper. Apart from this, the data
appeared essentially very clean, and just less than 600 seconds of data were removed on
the basis of either very high ($>30$\,ct s$^{-1}$) accepted event rates, very high
($>180$\,ct s$^{-1}$) master veto rates, or large values of atomic oxygen column
density along the line of sight.

Source detection and position determination were performed over the full field of
view with the EXSAS local detect, map detect, and maximum likelihood algorithms
(Zimmermann \etal\ \cite{Zimmermann92}). Images of pixel size 15\arcsec\ were
used for the source detection.

Sources accepted as detections were those with a likelihood L $\geq$10, and this
gave rise to a number of sources. Probabilities P, are related to maximum
likelihood values L, by the relation P$=1-e^{-\mbox{L}}$. Thus a likelihood L of
10 corresponds to a Gaussian significance of 4.0$\sigma$  (Cruddace \etal\
\cite{Cruddace}; Zimmermann \etal\ \cite{Zimmermann94}). 

Over the central 25\arcm$\times$25\arcm\ area (which we here concentrate on),
27 sources are detected, and these are listed in Table~\ref{table_srclist} as
follows: source number (col.\,1), corrected right ascension and declination
(cols.\,2,\,3), error on the source position (col.\,4, including a 3\farcs9
systematic attitude solution error), likelihood of existence (col.\,5), net
counts and error (col.\,6), count rates and errors after applying deadtime and
vignetting corrections (col.\,7), and 0.1$-$2.4\,keV flux, assuming a 5\,keV
thermal bremsstrahlung model (col.\,8). Count rates of the PSPC-detected point
sources can be converted into fluxes, assuming a variety of spectral models. A
1\,keV thermal bremsstrahlung model for instance, gives rise to fluxes 0.88
times those given in Table~\ref{table_srclist}. Two hardness ratios are given
in cols.\,9 \& 10, HR1, defined as (hard$-$soft)/(hard$+$soft) (hard and soft
being the net counts in the hard (channels 52$-$201) and soft (channels
8$-$41) bands, respectively), and HR2, defined as
(hard2$-$hard1)/(hard2$+$hard1) (hard1 and hard2 being the net counts in the
hard1 (channels 52$-$90) and hard2 (channels 91$-$201) bands, respectively).
The corresponding errors are also given (note that where a source is not
formally detected in one of the four bands, an upper or lower limit to the
hardness ratio has been calculated using a 2$\sigma$ upper limit to the
counts). HR1 is most sensitive to variations in the absorbing column, while
HR2 traces more the power law index or temperature. Hardness ratios can be
used to give very crude estimates of the spectral parameters that best
describe the source photons, and it may be worth comparing the tabulated
values with plots in the literature which show the variation of HR1 and HR2 for
simple spectral models (see \eg\ Pietsch \etal\ \cite{Pietsch98a}). The final
column (col.\,11) of Table~\ref{table_srclist} gives the nearest bright
optical counterpart from the APM finding charts of Irwin \etal\
(\cite{Irwin}); type (S$-$stellar, G$-$galaxy, F$-$faint), B magnitude, and
offset in arcseconds. Note that further correlations with several catalogues
were performed, notably with the SIMBAD database operated at CDS, Strasbourg.
Finally, the only source flagged as extended (and then, only at a likelihood
of 8.4) is source P27.

In an effort to improve the accuracy of the PSPC source positions, the RA and Dec
of five bright point sources (P3, P4, P12, P19 and P21) were compared with the
APM finding charts of Irwin \etal\ (\cite{Irwin}). Sources apparently associated
with the central galaxy, although they appeared to be very coincident, were not
used in the correction process, because of the confused and apparently 
extended nature of both
their optical and X-ray emission. A very small offset of 0\farcs04 in right
ascension and 0\farcs5 in declination is observed (the coordinates given in
Table~\ref{table_srclist} have been corrected for this).

\begin{table*}
\caption[]{X-ray properties of PSPC-detected point sources (see text).
Tabulated fluxes assume a 5\,keV thermal bremsstrahlung model and a hydrogen column
density of $N_{\rm H} = 1.14 \times 10^{20}$~cm$^{-2}$.}
\label{table_srclist}
\begin{tabular}{lrrrrrrrrrr}
\hline
\noalign{\smallskip}
Src.& \multicolumn{2}{c}{R.A.\,(J2000) Dec} & R$_{err}$ & Lik. & Net counts & Ct.\,rate &
$F_{\rm x}$ & \multicolumn{2}{c}{Hardness ratios} & Identification \\
 & ($^{\rm h~~~m~~~s}$) & ($^{\circ}~~~\arcm~~~\arcsec$) & ($\arcsec$) &  & &
(ks$^{-1}$) & ($\frac{10^{-14}{\rm erg}}{{\rm cm}^{2} {\rm s}}$) & (HR1) & (HR2) \\
\noalign{\medskip}
(1) & (2) & (3) & (4) & (5) & (6) & (7) & (8) & (9) & (10) & (11) \\
\noalign{\smallskip}
\hline
\noalign{\smallskip}
P1 &23 57 12.9&-32 23 57&16.6&  25.0& 72.1$\pm$12.8& 3.4$\pm$0.6 & 4.3$\pm$0.8& $>0.2$      &0.1$\pm$0.1 &F (24.1) 3\farcs0 \\
P2 &23 58 28.8&-32 24 42&17.3&  10.2& 33.1$\pm$9.3 & 1.6$\pm$0.5 & 2.0$\pm$0.6& $>0.1$      &0.8$\pm$0.3 &S (20.8) 27\farcs7 \\
P3 &23 58 42.4&-32 26 09&10.4&  63.3&107.7$\pm$13.8& 5.2$\pm$0.7 & 6.6$\pm$0.8&-0.2$\pm$0.1 &0.2$\pm$0.2 &S (19.5) 5\farcs0 \\
P4 &23 57 53.1&-32 28 10& 5.4& 279.4&219.0$\pm$16.7& 9.9$\pm$0.8 &12.5$\pm$1.0& 0.1$\pm$0.1 &0.1$\pm$0.1 &S (21.3) 2\farcs6 \\
P5 &23 57 44.3&-32 28 46&10.0&  26.7& 51.6$\pm$10.1& 2.3$\pm$0.5 & 2.9$\pm$0.6&-0.1$\pm$0.2 &$>0.5$      &F (23.0) 5\farcs6 \\
P6 &23 57 48.5&-32 32 30&15.0&  16.1& 41.3$\pm$9.6 & 1.8$\pm$0.4 & 2.3$\pm$0.5& 0.1$\pm$0.2 &$>0.1$      &S (16.1) 2\farcs8(?) \\
P7 &23 57 52.6&-32 33 11& 9.9&  31.9& 54.7$\pm$10.1& 2.4$\pm$0.5 & 3.0$\pm$0.6& 0.2$\pm$0.2 &0.2$\pm$0.2 &S (17.9) 1\farcs5(?) \\
P8 &23 57 59.8&-32 33 24&13.5&  29.5& 69.5$\pm$12.0& 3.1$\pm$0.5 & 3.9$\pm$0.7& 0.1$\pm$0.2 &0.0$\pm$0.2 &(?) \\
P9 &23 58 08.6&-32 34 03& 6.1& 111.7&110.1$\pm$12.5& 4.9$\pm$0.6 & 6.2$\pm$0.7& $>0.5$      &0.3$\pm$0.1 &F (21.2) 7\farcs3(?) \\
P10&23 57 47.1&-32 36 05&11.7&  79.8&173.4$\pm$18.1& 7.7$\pm$0.8 & 9.7$\pm$1.0& 0.1$\pm$0.1 &0.0$\pm$0.1 &NGC~7793 \\
P11&23 58 03.1&-32 36 34&12.9&  18.8& 41.2$\pm$9.3 & 1.8$\pm$0.4 & 2.3$\pm$0.5& $>0.2$      &0.4$\pm$0.2 &NGC~7793 \\
P12&23 58 38.6&-32 37 10& 6.2& 182.8&180.5$\pm$15.7& 8.4$\pm$0.7 &10.6$\pm$0.9& 0.0$\pm$0.1 &0.0$\pm$0.1 &G (18.3) 1\farcs7 \\
P13&23 57 51.2&-32 37 23& 4.1&2746.6&976.0$\pm$32.0&43.5$\pm$1.4 &54.9$\pm$1.8& 0.9$\pm$0.0 &0.4$\pm$0.0 &NGC~7793 \\
P14&23 57 31.5&-32 37 25&11.6&  17.9& 42.1$\pm$9.5 & 1.9$\pm$0.4 & 2.4$\pm$0.5& $>-0.5$     &$>0.1$      &S (20.3) 27\farcs6 \\
P15&23 58 32.6&-32 38 16& 9.0&  37.3& 62.1$\pm$10.4& 2.9$\pm$0.5 & 3.7$\pm$0.6& $>0.6$      &0.2$\pm$0.2 &F (20.9) 1\farcs1 \\
P16&23 57 12.0&-32 38 22&21.9&  14.5& 73.7$\pm$14.1& 3.4$\pm$0.6 & 4.3$\pm$0.8& 0.3$\pm$0.2 &0.3$\pm$0.2 &G (18.3) 3\farcs6 \\
P17&23 57 17.8&-32 39 50&12.9&  22.1& 52.9$\pm$10.6& 2.4$\pm$0.5 & 3.0$\pm$0.6&-0.1$\pm$0.2 &0.1$\pm$0.2 &F (21.2) 16\farcs1 \\
P18&23 57 25.3&-32 42 28&24.3&  15.5& 72.9$\pm$14.4& 3.3$\pm$0.7 & 4.2$\pm$0.8&-0.2$\pm$0.2 &0.0$\pm$0.3 &G (12.6) 14\farcs1 \\
P19&23 58 50.0&-32 42 45&10.6&  52.7& 99.1$\pm$13.5& 4.8$\pm$0.7 & 6.1$\pm$0.8& 0.1$\pm$0.2 &0.1$\pm$0.2 &S (19.3) 3\farcs5 \\
P20&23 58 28.5&-32 42 50&16.0&  14.5& 42.5$\pm$10.2& 2.0$\pm$0.5 & 2.5$\pm$0.6&-0.6$\pm$0.3 &$<1.0$      &F (20.9) 12\farcs3 \\
P21&23 57 51.4&-32 42 53&11.8&  23.9& 49.7$\pm$10.0& 2.3$\pm$0.5 & 2.9$\pm$0.6& 0.1$\pm$0.2 &0.2$\pm$0.2 &G (19.1) 5\farcs7 \\
P22&23 56 57.3&-32 43 32&19.8&  33.4&130.9$\pm$18.1& 6.3$\pm$0.9 & 8.0$\pm$1.1& 0.4$\pm$0.2 &0.2$\pm$0.1 &G (13.9) 4\farcs6 \\
P23&23 57 49.9&-32 45 48&11.3&  29.7& 60.5$\pm$10.9& 2.8$\pm$0.5 & 3.5$\pm$0.6& $<-0.2$     &$>-0.8$     &F (21.7) 0\farcs5 \\
P24&23 56 57.5&-32 46 34&38.0&  11.0& 66.9$\pm$15.7& 3.3$\pm$0.8 & 4.2$\pm$1.0&-0.2$\pm$0.2 &$>0.0$      &F (22.3) 5\farcs1 \\
P25&23 57 41.8&-32 46 53& 8.3&  94.9&131.6$\pm$14.4& 6.2$\pm$0.7 & 7.8$\pm$0.9& 0.0$\pm$0.1 &0.3$\pm$0.2 &F (21.2) 6\farcs2 \\
P26&23 57 14.0&-32 47 11&16.1&  61.0&164.0$\pm$18.7& 7.9$\pm$0.9 &10.0$\pm$1.1&-0.2$\pm$0.1 &0.3$\pm$0.2 &S (19.4) 11\farcs2 \\
P27&23 58 21.2&-32 47 17& 7.5& 249.8&302.3$\pm$20.9&14.5$\pm$1.0 &18.3$\pm$1.3&-0.4$\pm$0.1 &0.1$\pm$0.1 &S (18.8) 3\farcs7 \\
\noalign{\smallskip}
\hline
\end{tabular}
\end{table*}

Fig.\,1 shows a broad band (channels 8$-$235, corresponding approximately to
0.08$-$2.35\,keV) contour image of the central 25\arcm$\times$25\arcm\ region. All of the
sources listed in Table~\ref{table_srclist} are marked. Also shown in Fig.\,1 are three
smaller images, showing the very central (10\arcm$\times$10\arcm) emission, selected
over three separate spectral bands $-$ the `soft' band (channels 8$-$41), the `hard~1'
band (channels 52$-$90) and the `hard~2' band (channels 91$-$201).

\begin{figure*}
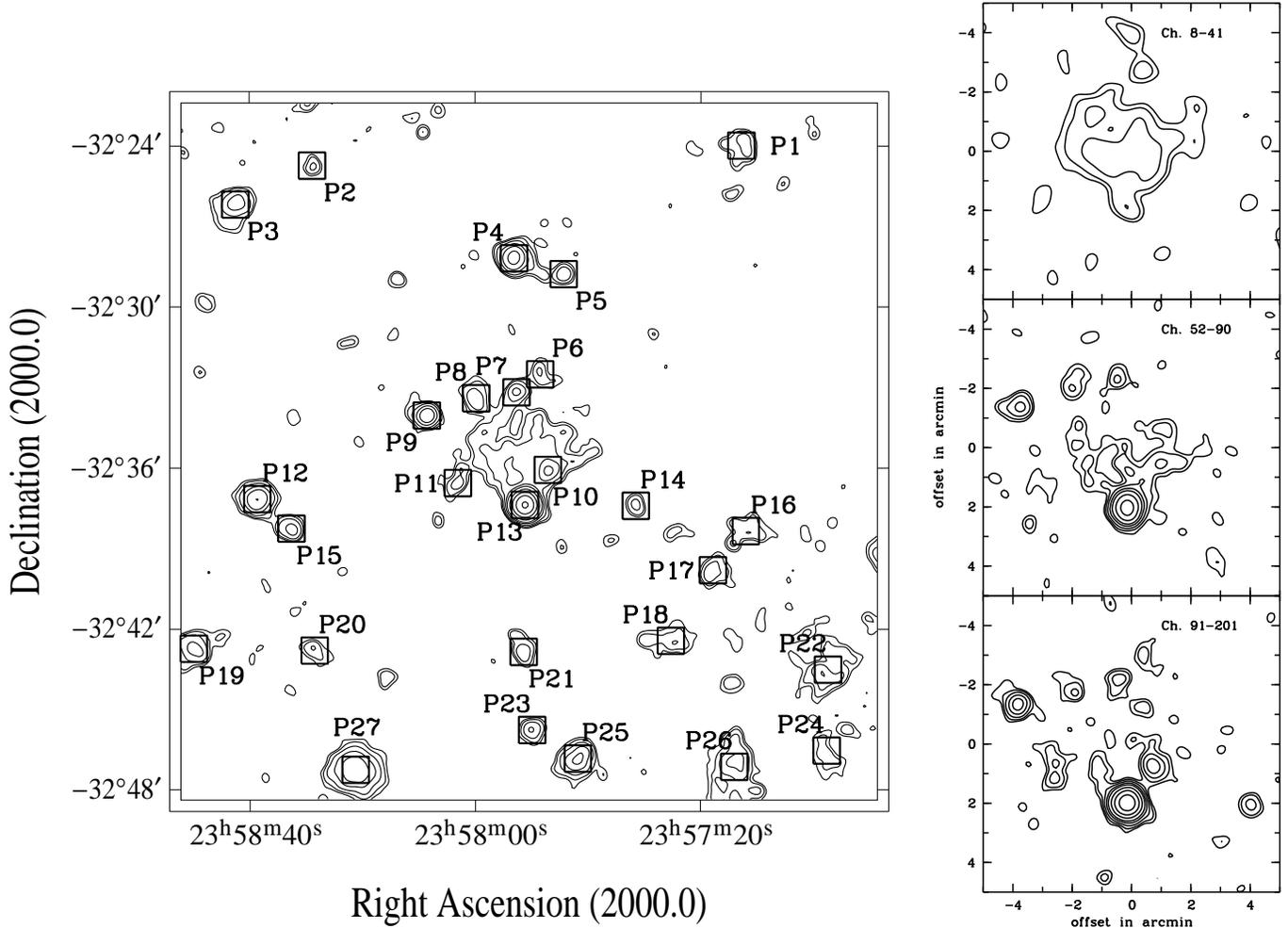

\unitlength1.0cm 
\begin{picture}(12.5,13.5)
\psfig{figure=7951.f1a,width=12.5cm,clip=}
\end{picture} 
\begin{picture}(5,13.5)
\psfig{figure=7951.f1b,width=5cm,clip=}
\end{picture} 
\hfill \parbox[b]{18.0cm} 
{\caption{\Ros\ PSPC maps of the NGC~7793 field in the broad (channels 8$-$235, corresponding
approximately to 0.08$-$2.35\,keV) band (main picture) and in the soft (channels 8$-$41),
hard~1 (channels 52$-$90) and hard~2 (channels 91$-$201) bands (three smaller pictures). The
contour levels in the broad- and soft-band figures are at 2, 3, 5, 9, 15, 31, 63, 127, 255,
511, 1023, 2047 and 4095$\sigma$ ($\sigma$ being $4.0\times10^{-4}$ (broad) and
$1.9\times10^{-4}$ (soft) cts s$^{-1}$ arcmin$^{-2}$) above the background
($1.7\times10^{-3}$ (broad) and $1.3\times10^{-3}$ (soft) cts s$^{-1}$ arcmin$^{-2}$). In
both the non-background-limited hard1- and hard2-band images, the contour levels are at 2, 3,
5, 9, 15, 31, 63, 127 and 255 times a value of $2.9\times10^{-4}$ cts s$^{-1}$ arcmin$^{-2}$.
Source positions, as given in Table~\ref{table_srclist}, are marked on the broad band image.
}}
\end{figure*}

To investigate whether any residual, low surface brightness emission is visible
within or around NGC~7793, an adaptive filtering technique was used to create an
intensity-dependent smoothed image. The signal-to-noise level of low surface
brightness emission is boosted by smoothing lower and lower intensity sections of
the image using Gaussians of progressively larger FWHM. Photons (from channels
8$-$235) were binned into an image of pixel size 8\arcsec. Pixels of amplitude 1
(2,3,4,5,6,7,8) were smoothed with a Gaussian of  FWHM 170\arcsec\
(120,85,60,40,30,20,15)\arcsec. Pixels of amplitude greater than 8 remained
unsmoothed, thus ensuring that the bright point sources were not smoothed into the
background regions. The resultant image is shown as a contour map overlaid on an
optical image in Fig.\,2 (the optical image is taken from the U.K.\,Schmidt plate
digitised sky survey).

\begin{figure*}
  \resizebox{12cm}{!}{
    \psfig{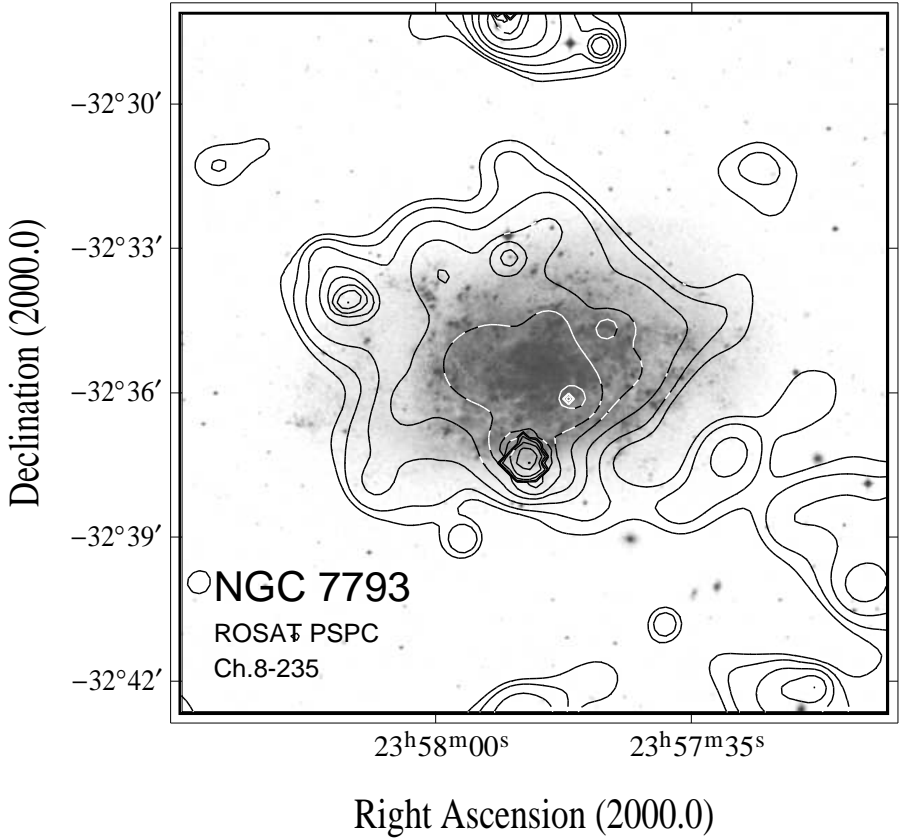}}
\hfill \parbox[b]{5.5cm} 
{\caption{\Ros\ (0.1$-$2.4\,keV) PSPC map of the central NGC~7793 field obtained using an
adaptive filtering technique (see text), overlaid on an optical image. The contour
levels are at 2, 3, 5, 9, 15, 31, 63, 127, 255, 511, 1023 and 2047$\sigma$ ($\sigma$
being $8.8\times10^{-5}$\,cts s$^{-1}$ arcmin$^{-2}$) above the background
($1.6\times10^{-3}$\,cts s$^{-1}$ arcmin$^{-2}$).
}}
\end{figure*}

The X-ray data was also compared with the NRAO VLA (NVSS) Sky Survey (Condon \etal 
\cite{Condon}), a radio continuum survey at 1.4\,GHz, covering the sky north
of -40$^{\circ}$ declination. Cross-correlating the NVSS catalogue with the 
sources listed in Table~\ref{table_srclist}) gave rise to radio counterparts lying 
within 20\arcsec\ of three X-ray sources, namely sources P14, P15 and P19. This
information was used to create Fig.\,3, where contours of broad-band 
X-ray emission (exactly as in Fig.\,1) are shown superimposed on the NVSS 1.4\,GHz 
continuum radio emission. 

\begin{figure*}
  \resizebox{12cm}{!}{
    \psfig{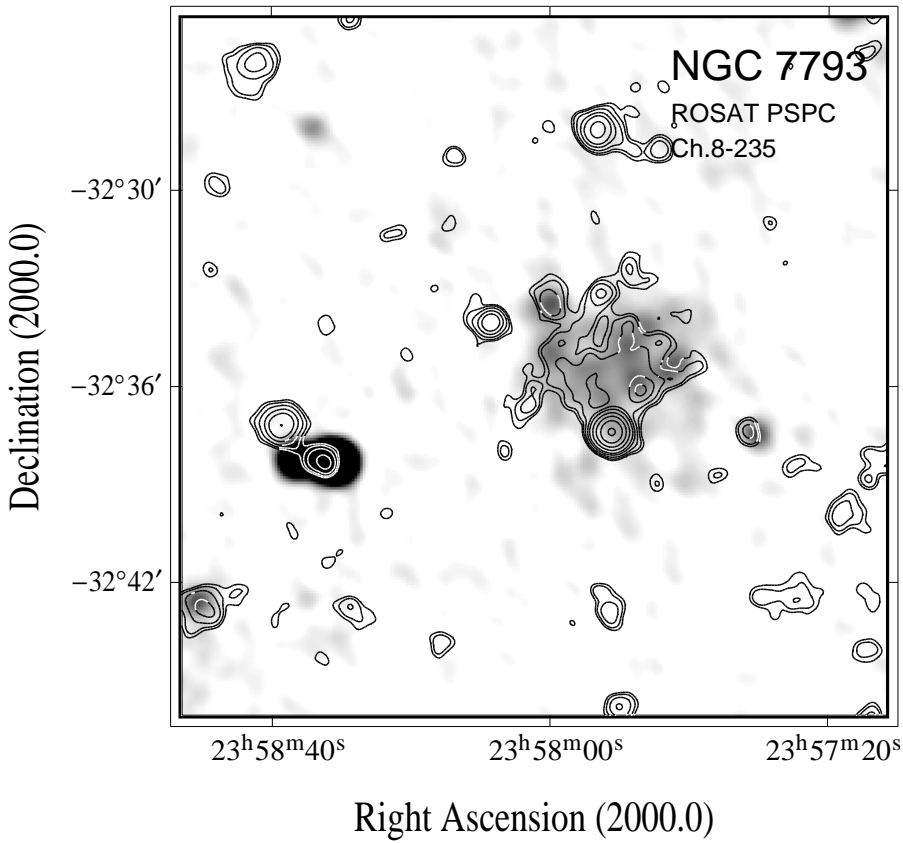}}
\hfill \parbox[b]{5.5cm} 
{\caption{The \Ros\ PSPC broad-band (channels 8$-$235) image, contoured as in 
Fig.\,1(main), and overlaid upon a NRAO VLA (NVSS) 1.4\,GHz 
radio continuum map (see text). 
}}
\end{figure*}

\subsection{Time variability study of point sources}
\label{timing}

A time variability study was performed on all of the central 27 PSPC-detected
sources.

As mentioned briefly earlier, essentially two separate observations were carried
out, almost half a year apart. The first of these (7/12/92) consisted of eight
observation blocks (of between 1.5 and 2\,ks). The second observation consisted
of six observation blocks (again, of between 1.5 and 2\,ks), the first five from
20/5/93, the last from 27/5/93. 

It was decided at first to bin the complete observation into these 2 main
observations separated by six months, the December 1992 and the May 1993
observation. A maximum likelihood search at the source positions given in
Table~\ref{table_srclist} was performed for both observation intervals, the
vignetting and deadtime corrected count rates (and errors) calculated within a cut
radius of $1.5\times$ the PSF FWHM at the source positions.
Table~\ref{table_halfyear} gives the count rates for the 27 sources for the two
observation intervals plus the probability that the source is variable. As can be
seen, a few of the sources appear variable, notably P9, P13 and P25. Two further
sources that appear to be variable at greater than the 2$\sigma$ significance level,
are P21 ($\sigma=2.4$) and P26 ($\sigma=2.5$).

\begin{table}
\caption[]{X-ray count rates of the 27 PSPC-detected point sources in the December 1992
observation and the May 1993 observation (see text).}
\label{table_halfyear}
\begin{tabular}{lrrr}
\hline
\noalign{\smallskip}
Source & \multicolumn{2}{c}{Count rate ($10^{-3}$\,s$^{-1}$)} & Prob.(var) \\
       & Dec.\,1992 & May 1993 &                                     \\
\noalign{\medskip}
(1) & (2) & (3) & (4) \\
\noalign{\smallskip}
\hline
\noalign{\smallskip}
P1  & 2.5$\pm$0.7 & 3.6$\pm$0.8 & 67\% \\
P2  & 1.4$\pm$0.6 & 2.0$\pm$0.7 & 48\% \\
P3  & 5.8$\pm$0.9 & 4.2$\pm$0.9 & 79\% \\
P4  &10.5$\pm$1.1 & 8.5$\pm$1.0 & 82\% \\
P5  & 2.4$\pm$0.6 & 1.9$\pm$0.6 & 41\% \\
P6  & 1.7$\pm$0.6 & 1.3$\pm$0.6 & 41\% \\
P7  & 2.6$\pm$0.6 & 1.7$\pm$0.5 & 37\% \\
P8  & 2.3$\pm$0.6 & 2.2$\pm$0.6 & 13\% \\
P9  & 8.0$\pm$0.9 & 0.8$\pm$0.4 &100\% \\
P10 & 5.1$\pm$0.8 & 5.2$\pm$0.9 &  7\% \\
P11 & 1.6$\pm$0.6 & 2.0$\pm$0.6 & 37\% \\
P12 & 8.6$\pm$1.0 & 7.3$\pm$1.0 & 63\% \\
P13 &36.9$\pm$1.7 &52.9$\pm$2.4 &100\% \\
P14 & 2.0$\pm$0.6 & 1.3$\pm$0.5 & 60\% \\
P15 & 3.4$\pm$0.7 & 2.2$\pm$0.6 & 79\% \\
P16 & 2.0$\pm$0.6 & 1.7$\pm$0.6 & 30\% \\
P17 & 2.6$\pm$0.7 & 2.1$\pm$0.6 & 42\% \\
P18 & 1.5$\pm$0.6 & 2.5$\pm$0.7 & 73\% \\
P19 & 4.6$\pm$0.9 & 4.8$\pm$0.9 &  8\% \\
P20 & 1.7$\pm$0.6 & 2.1$\pm$0.7 & 32\% \\
P21 & 3.0$\pm$0.7 & 1.1$\pm$0.5 & 98\% \\
P22 & 3.7$\pm$0.8 & 4.5$\pm$1.0 & 48\% \\
P23 & 2.6$\pm$0.7 & 2.8$\pm$0.7 & 13\% \\
P24 & 2.0$\pm$0.8 & 2.8$\pm$0.9 & 47\% \\
P25 & 7.6$\pm$1.0 & 3.5$\pm$0.8 &100\% \\
P26 & 6.9$\pm$1.1 & 3.6$\pm$0.9 & 99\% \\
P27 &13.8$\pm$1.3 &12.1$\pm$1.3 & 64\% \\
\noalign{\smallskip}
\hline
\end{tabular}
\end{table}

In an effort to search deeper into any further time variability present, the complete
observation was binned into the 14 convenient observation blocks described above, and
the same procedure was applied. Where a source was not detected with a likelihood
${\rm L} > 3.1$ (corresponding to a Gaussian significance of $2\sigma$), a $2\sigma$
upper limit to the count rate was calculated.

Fig.\,4 shows the results of this analysis for 14 of the 27 sources. These 14
sources are the sources apparently associated with NGC~7793 (sources P6, P7, P8,
P9, P10, P11 and P13) and those remaining sources containing in excess of 100 net
counts (sources P3, P4, P12, P22, P25, P26 and P27). The lightcurves of the
remaining 13 sources are generally of poor quality, and none show any significant
signs of variability. When viewing Fig.\,4, bear in mind that the x-axis is not
linear; timeslot (or observation block number) is plotted, not time. Almost half
a year exists between timeslots 8 and 9, and a week between timeslots 13 and 14.

Note that we are here interested in {\em variations} in the count rate from these
sources, and not so much in the absolute values of these count rates, as these have
been calculated earlier (Table~\ref{table_srclist}). Hence we have been able to use
smaller cut radii to avoid contamination from neighbouring sources. Differences in
the calculated count rates between these two methods, however, appear to be
negligible, except in cases involving sources embedded in extended emission (\eg\
P10).


\begin{figure*}
  \resizebox{12cm}{!}{
    \psfig{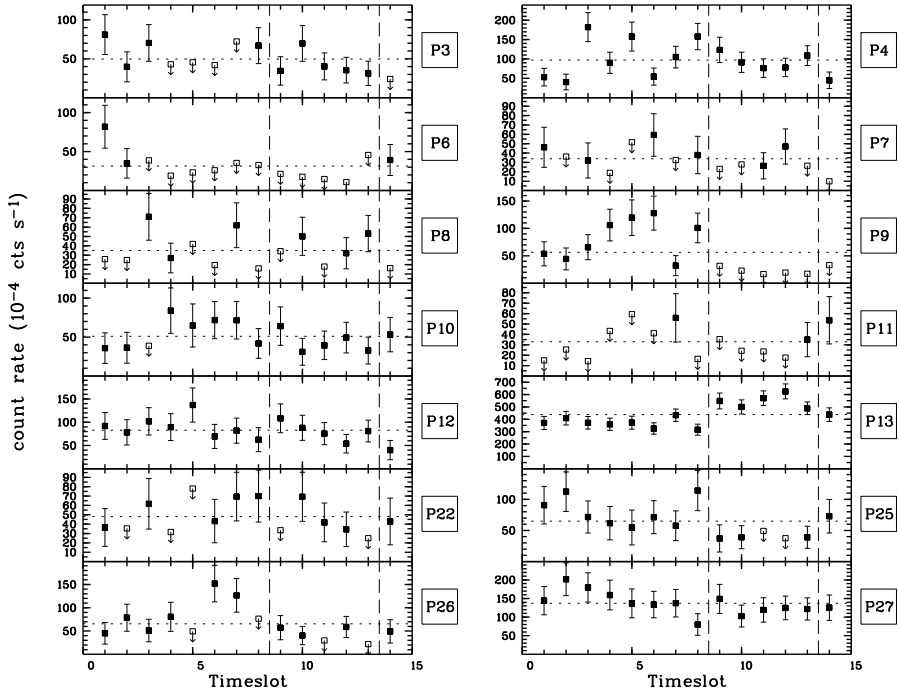}}
\hfill \parbox[b]{5.5cm} 
{\caption{\ros\ lightcurves of the seven detected point sources apparently associated with
NGC~7793, and the seven remaining bright ($>$100 net counts) sources.
Count rates are shown by filled squares with 1$\sigma$ error bars, and 2$\sigma$ upper
limits are shown by open squares. Dashed lines indicate the average count rate
calculated over the complete observation. Bear in mind that the x-axis is not linear in
time; half a year exists between timeslots 8 and 9, and a week between slots 13 and 14.
}}
\end{figure*}

\section{Results and discussion - the point sources}
\label{sec_disc_sources}

\subsection{The NGC~7793 point sources}

As can be seen in Fig.\,1, several sources are detected within the central region
of the PSPC field of view. Of particular interest to us are the sources
apparently associated with NGC~7793. These seven sources are indicated in
Table~\ref{table_srclist} by either a `NGC~7793' or a `(?)' in the last column,
and it is these sources that we shall discuss first.

Spectra of all seven point sources lying within the optical confines of
NGC~7793 (P6, P7, P8, P9, P10, P11 and P13) were analysed. Apart from the
very bright P13, spectra were extracted for the remaining six objects from
within circles of radius 1\farcm0. The P13 spectrum was extracted from a
circle of radius 1\farcm5.

A `pure' background spectrum was extracted from an annulus 5\arcm$-$7\farcm5 
from the centre of the field of view, with the data associated with each of
the intervening point sources having been excluded to a radius of 2\arcm. 

It is obvious from Fig.\,1, that several of the apparently NGC~7793-associated sources
lie above, or appear embedded within, unresolved, perhaps diffuse, emission. Because of
this, it was thought sensible to perform the source spectral fitting with not only the
`pure' background (extracted as explained above) as the background, but also with a
`background$+$unresolved' spectrum as the background. Though we describe the spectral
properties of the unresolved emission in the next section, the extraction of the spectrum
is described here, as it is used as a background spectrum in the fitting of the
NGC~7793-associated sources.

The unresolved spectrum (which obviously also contains `pure' background) was
extracted from a circle of radius 5\arcm, centred on the unresolved emission. 
Point source emission was excluded from this spectrum to radii of
0\farcm75 (1\farcm0 in the case of P13). 

The seven source spectra were background-subtracted (with both the `pure'
background and the `pure$+$unresolved' background), and, once corrected for
exposure and vignetting effects, were fitted with standard spectral models
(power law, thermal bremsstrahlung, blackbody and Raymond \& Smith
(\cite{Raymond}) hot plasma models). The results of all the best fits, both
using the `pure' background, and, in cases where they are of comparable or
better quality, those using the `unresolved' background, are given in
Table~\ref{table_fits_sources} as follows: Source (col.\,1) and spectral
model, whether PL - power law plus absorption, TB - thermal bremsstrahlung plus
absorption or BB - blackbody plus absorption (col.\,2) (note that Raymond \&
Smith hot plasma plus absorption models did not give a best fit in any of the
cases). Note also that a `U' in brackets indicates that the
`pure$+$unresolved' background has been used, as opposed to the `pure'
background. The next columns give the fitted $N_{\rm H}$ (col.\,3), the fitted
photon index $\Gamma$, where $F\propto E^{-\Gamma}$ (col.\,4), the fitted
temperature (kT, in keV) (col.\,5), and the reduced
$\chi^{2}$ (col.\,6). Note that in several cases (P7, P8 \& P10), very low 
values of reduced $\chi^{2}$ are obtained. Here the errors are so 
large that a large range of models is able to fit the data very well. 
Nothing really can be said here as regards whether one model is 
significantly better than another. 
Two values of the (0.1$-$2.4\,keV) luminosity are given
(cols.\,7\& 8), one (col.\,7) giving the `intrinsic' luminosity of the source
(\ie correcting for the total (\ie\ fitted) $N_{\rm H}$), the second (col.\,8)
giving an `emitted' luminosity (\ie correcting merely for the Galactic $N_{\rm
H}$; 1.14$\times10^{20}$\,cm$^{-2}$). All luminosities are calculated for an
assumed distance of 3.38\,Mpc.

Source P6 is X-ray faint, and lies some way from the optical confines of 
NGC~7793. It is however, almost coincident with a 16th magnitude stellar-like
object, less than 3\arcsec\ distant. P6's spectrum, though rather unconstrained, 
is consistent with that of quasars, which typically have power-law spectra with
photon indices in the range 2.2$\pm$0.2 (Branduardi-Raymont \etal\ 
\cite{Branduardi}; Roche \etal\ \cite{Roche}). P6 is therefore likely to be a 
background source.

Source P7 is in many ways similar to P6. Again X-ray faint, and lying some way
from the centre of NGC~7793, it too has a stellar-like (18th magnitude)
counterpart, less than 2\arcsec\ distant, visible in the H$\alpha$ + \nii\ and
the \sii\ images of Blair \& Long
(\cite{Blair}), as a strong, slightly fuzzy source, which the authors do not
attribute to anything belonging to NGC~7793. Again, as in the case of P6, P7's
spectrum is consistent with it being due to a background QSO (though the photon
index is a little on the low side). Interestingly, a feature quite 
close to source P7 is visible in the \Ein\ IPC image (Fabbiano \etal\
\cite{Fabbiano92}) (it actually lies north of P7 and east of P6), though only a handful
of counts were received. Again, like P6, P7 may be unrelated to NGC~7793, and instead
is likely due to a background object.

Source P8 is very interesting, as it appears coincident (to within 3\arcsec) of a
very bright H$\alpha$ and \sii\ complex, identified in Blair \& Long (\cite{Blair})
as being associated with a supernova remnant (their S26), one of the largest and
brightest in their NGC~7793 sample. The fact that the X-ray spectrum of P8 is best
fit by a low-temperature thermal model is very encouraging, and adds credence to the
idea that S8 is associated with the supernova remnant S26. Furthermore, note that an
enhancement in 1.4\,GHz radio continuum emission is seen coincident with P8
(Fig.\,3), as would be expected if supernova activity were ongoing at this position.
In fact, the situation is even more interesting than this as, as discussed in Blair
\& Long \cite{Blair}, the SNR NGC~7793-S26 is only the bright, high surface
brightness portion of a much more extended, high [\sii]:H$\alpha$ ratio, roughly
oval emission region, which is thought, on account of its appearance, size, surface
brightness, and lack of multiple loop structures, to be a collection of multiple
supernovae, blowing out a large bubble of shock-excited gas, perhaps eventually
becoming a superbubble. It is useful to compare the spectral properties of known
\hii\ regions and superbubbles with P8. Williams \& Chu (\cite{Williams}) present
hardness ratios (calculated in their paper as the difference divided by the sum of high-
(1.0$-$2.5\,keV) and low- (0.1$-$1.0\,keV) counts) for \hii\ regions in M101 and
superbubbles in the LMC. The hardness ratio (calculated in the same way) for P8
($\approx -$0.52) is soft, and agress very favourably both with the majority of the
M101 \hii\ regions, and all of the Williams \& Chu (\cite{Williams}) superbubbles.
We must be careful however, as it can be seen in Fig.\,4 that there may be some
evidence for time-variability within source P8. If this were true, then this would
point against P8 being due to a superbubble. Note that the limits in Fig.\,4 are not
very restrictive however, and that the probability of time-variability for P8, when
comparing the two main observations, is very low (see Table~\ref{table_halfyear}).

Source P9, though quite bright, appears coincident with nothing whatsoever. Situated
at the very north-eastern edge of a filament of material from NGC~7793, it may well
be associated with the host galaxy, though no bright \hii\ regions or plausible
sites of significant supernova activity lie anywhere within 1\arcm. Also, no
enhancement at all in the radio emission is seen at this position (see Fig.\,3),
and the nearest optical counterpart, over 7\arcsec\ distant, is very faint (below
21st magnitude). The X-ray spectrum of P9, though very unconstrained, is very
strange also, being very steep, but very absorbed. One other interesting facet of
P9's X-ray emission can be gleaned from Fig.\,4 and the discussion within
Section~\ref{timing} $-$ it is most definitely variable. There is a very significant
difference between the X-ray strength of P9 during the first observation (Dec.\,92),
where some evidence for a gradual rise in brightness is seen, and the strength of
the emission during the second observation (May 93), where very little emission is
seen (see Table~\ref{table_halfyear}). This change in the flux level of source P9 is
very apparent in Fig.\,5, where contours of hard band (channels 52$-$201) X-ray
emission are shown for each of the two separate observations, the December 1992
observations and the May 1993 observation. In the \Ein\ image of Fabbiano \etal\
(\cite{Fabbiano92}), a feature is observed less than 15\arcsec\ from the position of
P9. Again however, as with the \Ein\ feature tentatively associated with P7/P6, very
few counts (around 6$-$7) were obtained. P9 is probably though, on account of its
lying within the optical confines of NGC~7793, and its unusual spectrum, associated
with NGC~7793.

\begin{table*}
\caption[]{Results of the best model fits to the seven possibly
NGC~7793-associated source spectra (see text). Models are: PL (power law
plus absorption), TB (thermal bremsstrahlung plus absorption), 
BB (blackbody plus absorption). A bracketed `U' indicates that the `unresolved' 
background (as opposed to the `pure' background) has been used. Errors on the 
spectral parameters are 1$\sigma$. 
Two (0.1$-$2.4\,keV) luminosities are tabulated.
One, the intrinsic PSPC luminosity of the source, and two, the Galactic
$N_{\rm H}$-corrected (\ie emitted) PSPC luminosity (Galactic $N_{\rm H} =
1.14 \times 10^{20}$~cm$^{-2}$). 
}
\label{table_fits_sources}
\begin{tabular}{llrrrrrr}
\hline
\noalign{\smallskip}
Source & Model & $N_{\rm H}$ & Photon & $kT$  & red.$\chi^{2}$ & 
\multicolumn{2}{c} {$L_{\rm x}$ (10$^{38}$\,erg s$^{-1}$)} \\ 
 & & 10$^{20}$\,cm$^{-2}$& Index    &(keV)& & (Intrinsic)& (Emitted) \\
(1) & (2) & (3) & (4) & (5) & (6) & (7) & (8) \\
\noalign{\smallskip}
\hline
\noalign{\smallskip}
P6  & PL  & 1.9$^{+5.3}_{-1.9}$ & 2.18$\pm$2.06 &         & 1.23 & 0.38$\pm$0.09 & 0.27$\pm$0.07 \\
    & TB  & 0.9$^{+2.7}_{-0.9}$ & & 1.24$^{+1.62}_{-1.24}$         & 1.23 & 0.26$\pm$0.06 & 0.26$\pm$0.06 \\

P7  & PL  & 1.9$\pm$1.1 & 1.37$\pm$1.59 & &         0.19 & 0.40$\pm$0.09 & 0.36$\pm$0.08 \\
    & TB  & 1.6$^{+2.7}_{-1.6}$ & & 12.9$^{+173}_{-12.9}$& 0.19 & 0.39$\pm$0.09 & 0.36$\pm$0.08 \\

P8  & TB  & 7.8$^{+8.9}_{-7.8}$ & & 0.36$\pm$0.25 &        0.21 & 1.38$\pm$0.30 & 0.39$\pm$0.08 \\
    & BB  & 3.1$^{+6.5}_{-3.1}$ & & 0.17$\pm$0.05 &         0.16 & 0.44$\pm$0.10 & 0.35$\pm$0.08 \\

P9  & PL  &37.3$\pm$8.9 & 3.91$\pm$4.18 & &         1.02 & 52.6$\pm$7.42 & 0.64$\pm$0.09 \\
    & TB  &20.9$^{+46.8}_{-20.9}$& &0.64$^{+0.94}_{-0.64}$ &  1.04 & 2.66$\pm$0.38 & 0.64$\pm$0.09 \\
   
P10 & TB  & 2.9$\pm$1.6 & & 0.77$\pm$0.64 &         0.14 & 1.24$\pm$0.14 & 0.83$\pm$0.09 \\
    &TB(U)& 3.3$\pm$2.2 & & 0.63$\pm$0.52 &         0.06 & 1.24$\pm$0.14 & 0.74$\pm$0.08 \\
 
P11 & PL  & 0.1$^{+3.8}_{-0.1}$ & 0.72$\pm$0.81 & &         1.84 & 0.31$\pm$0.07 & 0.31$\pm$0.07 \\
    &PL(U)& 1.3$^{+10.0}_{-1.3}$& 0.78$\pm$1.41& &  0.71 & 0.28$\pm$0.06 & 0.27$\pm$0.06 \\

P13 & PL  & 12.0$\pm$8.2& 1.76$\pm$0.45 & &         1.21 & 17.8$\pm$0.59 & 8.68$\pm$0.29 \\
    & TB  & 9.8$\pm$3.8 & & 3.49$^{+4.26}_{-3.49}$ &         1.27 & 14.0$\pm$0.46 & 8.71$\pm$0.29 \\
    &PL(U)& 13.5$\pm$10.9&1.76$\pm$0.55 & &         1.26 & 18.0$\pm$0.59 & 8.49$\pm$0.28 \\

\noalign{\smallskip}
\hline
\end{tabular}
\end{table*}

\begin{figure*}
\unitlength1.0cm 
\begin{picture}(8.9,8.9)
\psfig{figure=7951.f5a,width=8.9cm,clip=}
\end{picture} 
\begin{picture}(8.9,8.9)
\psfig{figure=7951.f5b,width=8.9cm,clip=}
\end{picture} 
\hfill \parbox[b]{18cm} 
{\caption{\Ros\ PSPC maps of the centre of the NGC~7793 field in the hard band
(channels 52$-$201, corresponding approximately to 0.52$-$2.01\,keV) for (left)
the December 1992 observation and (right) the May 1993 observation, superimposed 
on an optical image. The contour
levels in each figure are at 2, 3, 5, 9, 15, 31, 63 and 127 
times a value of $6.0\times10^{-3}$ cts s$^{-1}$ arcmin$^{-3}$. 
}}
\end{figure*}

Source P10 lies nearest to the centre of NGC~7793, and appears embedded in diffuse
emission (as an aside, it is interesting to note that no nuclear enhancement
is visible, either as a nuclear point source, or as an increase in unresolved
emission). Though in a region containing much supernova activity and \hii\
regions, no particular sources stick out as probable counterparts within the work
of Blair \& Long (\cite{Blair}). The same is true as regards the radio emission 
(Fig.\,3) $-$ though a number of enhancements in the radio emission are visible 
within the NGC~7793 disc, none are particularly coincident with P10.
P10 is likely however, to be due, as in the case 
of source P8, to a superbubble, or collection of supernova remnants, given the
fact that it appears constant in flux (Fig.\,4) and that a good thermal
low-temperature fit to its spectrum is obtained. 
A fitted absorbing column of greater than
(although consistent with) the Galactic column, may indicate that source P10
lies, in some way, beneath a layer of diffuse emission.

Source P11 appears similar to both source P6 and P7, in that it is a faint X-ray
source with a bright stellar-like counterpart that shows up both as a source of
H$\alpha$ and \sii\ emission (Blair \& Long \cite{Blair}). Also, as with sources P6
and P7, no significant radio counterpart appears to exist, and some hint of X-ray
variability is present, though the low number of counts (in all of these three
cases) make it impossible to attach any degree of certainty to this last statement.
Interestingly, a significantly better fit to P11's spectrum is obtained, when the
diffuse emission is included in the background (indicating that significant diffuse
emission is present at this position, as is evident from Fig.\,2). The best
power-law fit (both with `pure' and with `pure + unresolved' background) suggests,
in contrast to sources P6 and P7, a very flat spectrum, indicating a non-QSO origin.
Given this, and the optical correlations described above, P11 is likely to belong to
NGC~7793.

Source P13, the brightest source in the vicinity of NGC~7793 (and indeed in the
entire PSPC field of view), is perhaps the most interesting source detected. It
is clearly visible in the \Ein\ image of Fabbiano \etal\ (\cite{Fabbiano92}),
dominating the emission from NGC~7793. Though the X-ray position they quote
lies some 15\arcsec\ east and 28\arcsec\ north of P13, it should be noted that
the \Ein\ positions are not thought to be too accurate ($\sim$1\arcmin; \eg\
Margon \etal\ (\cite{Margon})). This 1850\,s \Ein\ observation obtained almost 40
source counts from within a circle of radius 2\farcm5, leading to an estimated flux
of 6.81$\times10^{-13}$\,erg cm$^{-2}$ s$^{-1}$. Note that in the \Ein\ case, (1)
this circle, apart from containing P13, also contains our sources P10 and P11, (2)
this circle will also contain a hard (\gtsim 0.5\,keV) component of the unresolved
emission, and (3), in calculating their flux value, Fabbiano \etal\
(\cite{Fabbiano92}) were only, due to the very poor spectral resolution of the \Ein\
IPC, able to assume a nominal 5\,keV thermal bremsstrahlung spectrum.

P13 has had a series of optical follow-up observations made on it. It appears for
instance, in Margon \etal's (\cite{Margon}) atlas of X-ray selected
quasi-stellar objects, in which optical identifications of objects
serendipitously detected by \Ein, were made. Here, spectra were obtained for
all candidate objects within the X-ray position regions, to a limit of
$B=18.5-19$. An object (named 2355-329) with a QSO-type spectrum is observed in
the vicinity of source P13, at a redshift (the smallest in Margon \etal's
(\cite{Margon}) sample) of 0.071 (corresponding to a distance of 280\,Mpc,
assuming H$_{0} = 75$\,km s$^{-1}$ Mpc$^{-1}$). The object is very strong in
[\oii] $\lambda 3727$, and shows a large UV excess on the CTIO 1\,m plate.

More recently, Bowen \etal\ (\cite{Bowen}) 
have presented finding charts and optical positions, accurate to $<$1\arcsec, for 
several QSOs, including the one to the south of NGC~7793. They name it 2355-3254
and give its 2000.0 position as 
$\alpha=$23$^{\rm h}$57$^{\rm m}$52.39$^{\rm s}$, 
$\delta=$-32$^{\circ}$38\arcm17.1\arcsec. 
Margon \etal\ (\cite{Margon}) quote the large discrepancy between the optical 
and \Ein\ positions (just over an arcminute), stating that the \Ein\ positions
are `accurate to of order 1\arcm'. 

Though this may have been true, the problem 
still exists $-$ the \Ros-detected source P13 lies just less than an arcminute
north-by-northwest of the optical source 2355-329. This is shown clearly in 
Fig.\,6, where contours of broad-band (Ch.\,8$-$235) X-ray emission are 
shown superimposed upon on optical image of the southern part of NGC~7793. 
The Margon \etal\ (\cite{Margon}) proposed optical counterpart, the QSO at a
redshift of 0.071, is actually the leftmost of the two point-like objects, an
arcminute or so to the south of the X-ray contour peak.

\begin{figure}
\unitlength1.0cm 
\begin{picture}(8.5,8.5)
\label{fig_spec_ic48.5}
\psfig{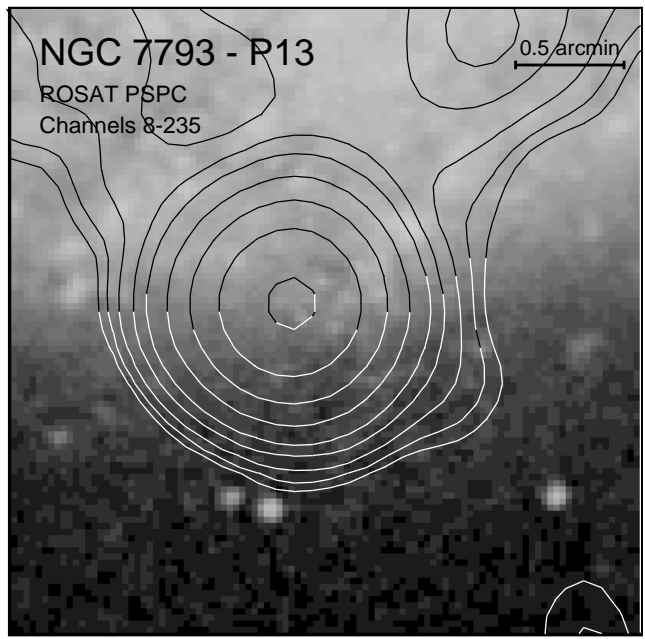}
\end{picture} 
\hfill \parbox[b]{8.7cm} 
{\caption{Contours of broad band (Ch.\,8$-$235) X-ray emission (as in Fig.\,1), 
superimposed on an optical image of the southern part of NGC~7793. 
The proposed optical QSO counterpart is the leftmost of the two point-like
objects, an arcminute or so to the south of the X-ray contour peak.
}}
\end{figure}

The \Ros\ X-ray position is far more accurate than the \Ein\ position (just over
4\arcsec, in the case of P13), and, as such, we can conclude that the bright X-ray
source P13 is {\em not} associated with the QSO-type object 2355-329. The spectral
properties of source P13 are consistent with that of a QSO or high-mass X-ray binary.
The best fit to the P13 spectrum (tabulated in Table~\ref{table_fits_sources}, and
shown in Fig.\,7) is a power-law model with a photon index of 1.76$\pm$0.45 (see above
discussion regarding P6).


P13 does appear to be significantly variable. While the fitting of the P13 lightcurve
(see Fig.\,4) to a constant flux level results in a $\chi^{2}$ of 40.85 (suggesting
a variability at the 3.9$\sigma$ ($L$=9.21) significance level), the difference in 
count rates obtained during the December 1992 and May 1993 observations (see
Table~\ref{table_halfyear}) confirm that P13 is most definitely variable. 

Given the lack of any obvious counterpart (as is evident both in Fig.\,6
and in Blair \& Long (\cite{Blair})), and the evidence presented here, the
true nature of P13 still remains a mystery. The variability is strongly
suggestive of the emission being due to a single object, and not, for
instance, to a collection of X-ray binaries, say. It could be due a
background QSO, though the emission spectrum is a little flat when compared
with typical QSOs, and it would have to be a radio quiet system, as no
significant enhancement in radio emission is seen (Fig.\,3). P13 could be a
young supernova, as these can attain X-ray luminosities equal to or greater
than that of P13. SN~1980K (in NGC~6946), SN~1986J (in NGC~891), SN~1978K
(in NGC~1313), SN~1993J (in M81) (see Schlegel (\cite{Schlegel95}) and the
references therein) and SN~1979C (in M100; Immler \etal\ \cite{Immler})
have all attained X-ray luminosities greater than or equal to that of P13. 
Furthermore, even though P13's emission is hard,
its hardness ratio (calculated as in the case of P8) being positive
($\approx$+0.14), supernovae are seen to show a large range in spectral
hardness (Schlegel \cite{Schlegel95}; Bregman \& Pildis (\cite{Bregman92}).
The fact, however, that nothing significant is seen in the radio
(Fig.\,3), and that the putative supernova was never actually observed in
outburst, suggest strongly against the supernova argument. If P13 is
contained within NGC~7793, then it is certainly a superluminous source. It
is unlikely to be an X-ray nova, unless we have caught it in outburst in
both the December 1992 and the May 1993 observations. Instead it is most
likely (if it is contained within NGC~7793) to be due to an accreting
binary. Bearing in mind however, that the Eddington limit for a $1M_\odot$
compact object is $1.3\times 10^{38}$\,erg~s$^{-1}$, then P13's X-ray
emission suggests the prescence of a massive ($\sim10M_\odot$) black hole
X-ray binary. 
Finally, nothing more conclusive can be said regarding the nature of P13,
when one considers its $f_{\rm X}/f_{\rm opt}$ ratio. One can estimate a
limiting optical magnitude at the position of P13 from the U.K.\,Schmidt
digitised sky survey optical image (shown in Figs.\,2, 5 and 6). This limit
($\gtsim19$), when used in conjunction with the method of Maccacaro \etal\
(\cite{Maccacaro88}), gives rise to a $f_{\rm X}/f_{\rm opt}$ ratio of around 
5, rather high, but consistent with that of AGNs and X-ray binaries.

\begin{figure*}
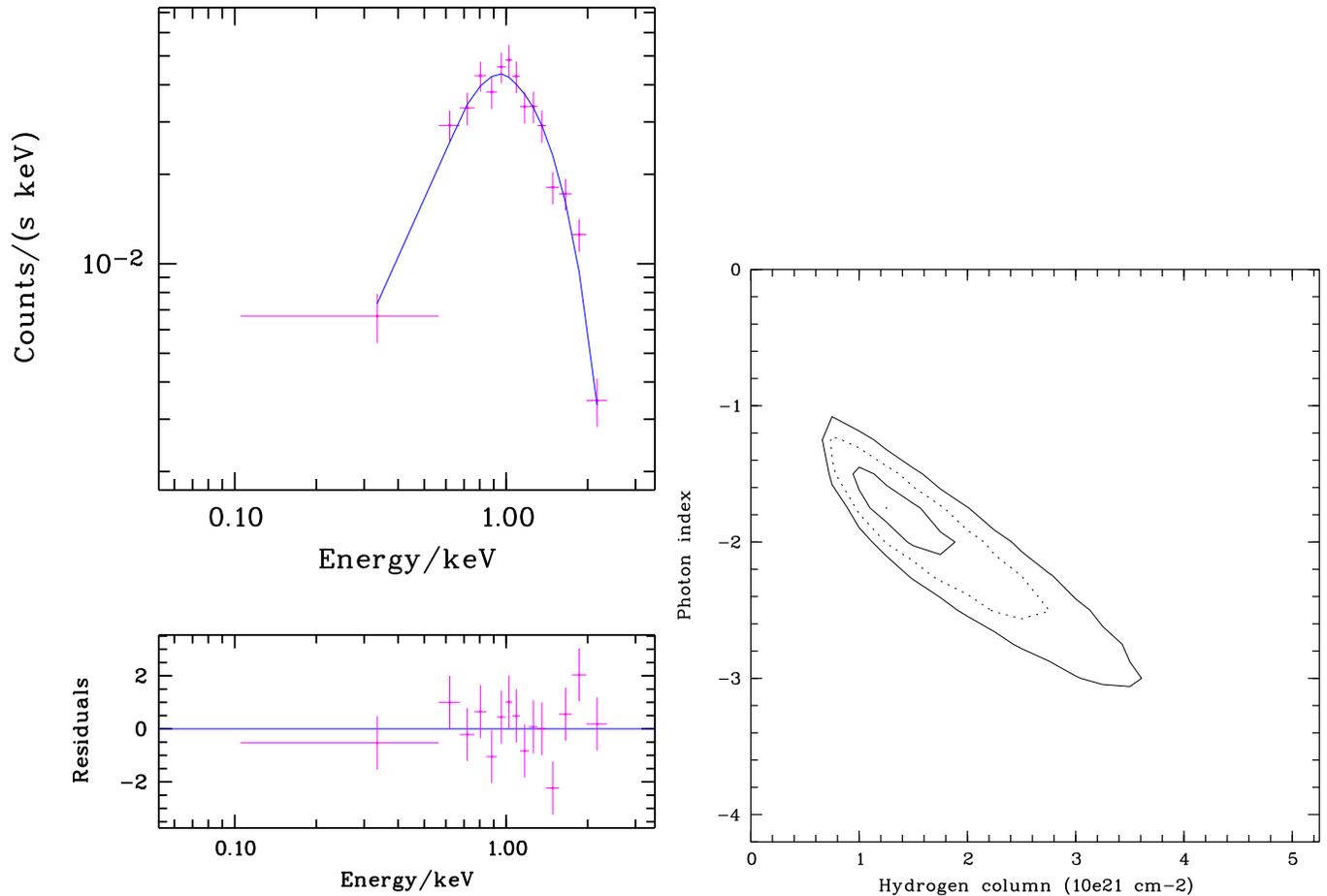

\unitlength1.0cm 
\begin{picture}(8.9,12.0)
\psfig{figure=7951.f7a,width=8.9cm,clip=}
\end{picture} 
\begin{picture}(8.9,8.9)
\psfig{figure=7951.f7b,width=8.9cm,clip=}
\end{picture} 
\hfill \parbox[b]{18cm} 
{\caption{ (Left) P13 spectrum with the best-fit power-law plus absorption model
(see Table~\ref{table_fits_sources}). The pulse height spectrum of the total X-ray 
emission is indicated by crosses, and the fit, by the solid line. 
(Right) Gaussian contour levels of
1$\sigma$, 2$\sigma$ and 3$\sigma$ in the photon index$-$hydrogen column plane for
the power-law fit to the P13 
spectrum. 
}}
\end{figure*}

\subsection{The secondary point sources}

Many of the remaining bright sources within the NGC~7793 field are interesting. P3,
appears to be non-variable, and is undoubtedly associated with a star-like object
($B$ magnitude = 19.5), given the good spatial correlation between the object
position and the X-ray position (5\arcsec), and the lack of any other, even faint
optical counterpart within more than half an arcminute. The identity of P4 is more
uncertain, as two very faint sources lie within 3\arcsec\ of the X-ray position. A
bright star ($B$=15.5), lying nearly 17\arcsec\ distant, is unlikely to be the
source, given the error in the X-ray source position (5\farcs4; see
Table~\ref{table_srclist}). P12 is almost definitely associated with a ($B$=18.3)
galaxy. The spatial correlation between optical and X-ray position is excellent
(1\farcs7), and no other, even faint feature is observable out to 25\arcsec. Note
also that P12 appears to be very constant in flux. Source P22 is, in many ways,
identical to P12, though the identification is less certain. Though a bright
($B$=13.9) galaxy, 4\farcs6 from the P22 X-ray position, is likely to be the host,
the error on the X-ray position is large (nearly 20\arcsec), encompassing a faint
($B$=20.2) star. Though the nearest feature to source P25 (as given in
Table~\ref{table_srclist}) lies some 6\farcs2 distant, the true optical counterpart
is perhaps more likely to be a far brighter ($B$=15.6) star-like feature, 12\arcsec\
distant (note though, that the error on the X-ray position is somewhat less than
this). The nearest source to P26's X-ray position (a $B$=19.4 star-like object) lies
some 11\farcs2 distant, a problem perhaps, if it weren't for the fact that the error
on P26's X-ray position (16\farcs1; Table~\ref{table_srclist}) is large. Finally,
P27 is almost certainly associated with a $B$=18.8 star-like object, nearly 4\arcsec
away. A small error in the X-ray position, together with a constant X-ray flux level,
and a lack of any other optical features within more than 20\arcsec\ of the X-ray
position, makes this statement quite secure.

P14, P15 and P19 all appear to have radio counterparts (see Fig.\,3 and the
discussion thereof). Though no optical counterpart to P14 is seen out to a distance
of almost 0\farcm5, P19 lies only 3\farcs5 from a ($B$=19.3) star-like object. Both
objects are likely to be background AGN. P15 seems to be the most interesting of the
three however, as it lies between two very bright radio features (which have been
overexposed in Fig.\,3). Furthermore, a very faint ($B$=20.9) optical feature is seen
1\farcs1 away from the X-ray position. This source is very suggestive of being an
X-ray bright AGN with a  double-lobed radio source (the rightmost lobe seemingly
being directed towards us). Many sources of this type have been found in the REX
survey (Maccacaro \etal\ \cite{Maccacaro98}), a project aiming, through the
cross-correlation of radio and X-ray surveys, to select large samples of BL~Lac
objects and other kinds of radio loud AGN. P15 appears to be a REX (a Radio Emitting
X-ray source), and may indeed be a BL~Lac. 


\section{Results and discussion - the unresolved emission}
\label{sec_disc_unres}

\subsection{The unresolved emission - spectral properties}

As has been discussed briefly before and, as is evident from both the soft band
image of Fig.\,1 and from Fig.\,2, a good amount of unresolved, perhaps diffuse
emission is present. Good evidence that a significant fraction of this unresolved
X-ray emission is truly diffuse is seen in the radio (Fig.\,3), where what appears
to be diffuse continuum emission is seen covering the NGC~7793 disc. This diffuse
radio emission is very likely due to supernovae and supernova remnants, which would
give rise to diffuse X-ray emitting gas also.

A spectrum of this unresolved X-ray emission, together with a background
spectrum were extracted, as described in Sect.~\ref{sec_disc_sources}. As in
the fitting of the point source spectra, spectral models were fitted to this
unresolved emission spectrum. Power law plus absorption, thermal
bremsstrahlung plus absorption and Raymond \& Smith hot plasma plus absorption
models were each used in the fitting process, and the results of the best
model fits are given in Table~\ref{table_fits_diff}, the columns being
identical to those of Table~\ref{table_fits_sources} (see
Sect.~\ref{sec_disc_sources}). Note again, that the very low values of 
reduced $\chi^{2}$ are really a reflection of the statistics, and nothing
concrete can be said regarding whether one model is significantly better 
than another. 
Note also that all luminosities (calculated
again for an assumed distance of 3.38\,Mpc) have been scaled up to allow for
the emission lost in the `holes' left after the source subtraction procedure.

\begin{table*}
\caption[]{Results of the best model fits to the unresolved emission spectrum (see
text). Models are: PL (power law plus absorption), TB (thermal bremsstrahlung
plus absorption), RS (Raymond \& Smith hot
plasma plus absorption). An `F' indicates that the parameter has been frozen at the 
value given. Errors on the spectral parameters are 1$\sigma$. 
Two (0.1$-$2.4\,keV) luminosities are tabulated. One,
the intrinsic PSPC luminosity of the emission, and two, the Galactic
$N_{\rm H}$-corrected (\ie emitted) PSPC luminosity (Galactic $N_{\rm H} = 1.14
\times 10^{20}$~cm$^{-2}$). Both assume a distance of 3.38\,Mpc.
}
\label{table_fits_diff}
\begin{tabular}{lrrrrrrr}
\hline
\noalign{\smallskip}
Model & $N_{\rm H}$ & Photon & $kT$  & $Z$ & red.$\chi^{2}$ & 
\multicolumn{2}{c} {$L_{\rm x}$ (10$^{38}$\,erg s$^{-1}$)} \\ 
 & 10$^{20}$\,cm$^{-2}$& Index    &(keV)& (Solar)& & (Intrinsic)& (Emitted) \\
(1) & (2) & (3) & (4) & (5) & (6) & (7) & (8) \\
\noalign{\smallskip}
\hline
\noalign{\smallskip}
%
RS & 0.9$^{+4.4}_{-0.9}$& & 1.01$\pm$0.56 & 0.06$^{+0.11}_{-0.06}$& 0.36 & 4.07$\pm$0.42 & 4.07$\pm$0.42 \\
   & 1.14(F)            & & 0.97$\pm$0.30 & 0.04$\pm$0.02         & 0.31 & 4.32$\pm$0.45 & 4.32$\pm$0.45 \\

TB & 1.8$\pm$1.8 & & 0.85$\pm$0.40 &               & 0.49 & 5.31$\pm$0.55 & 4.27$\pm$0.44 \\
   & 1.14(F)     & & 1.11$\pm$0.48 &               & 0.51 & 4.59$\pm$0.47 & 4.59$\pm$0.47 \\

PL & 3.5$\pm$3.4 & 2.58$\pm$1.03 & &               & 0.65 &10.53$\pm$1.09 & 4.47$\pm$0.46 \\
   & 1.14(F)     & 1.85$\pm$0.25 & &               & 0.93 & 4.80$\pm$0.50 & 4.80$\pm$0.50 \\


%
\noalign{\smallskip}
\hline
\end{tabular}
\end{table*}

As can be seen, thermal models give very good results, the Raymond \& Smith (RS) hot
plasma model being the best (though all are good). The unresolved emission  spectrum
and the best (RS) model is shown in Fig.\,8. The size of the error regions can be seen
in Fig.\,9, where Gaussian contour levels of 1$\sigma$, 2$\sigma$ and 3$\sigma$ are
shown in the metallicity$-$temperature and the temperature$-$absorption column plane
for the Raymond \& Smith hot plasma model (note that the equivalent bottom plot for
the thermal bremsstrahlung model is essentially identical, though the error regions are
slightly larger). The RS absorption column value is entirely consistent with the
column out of our own Galaxy in the direction of NGC~7793,
1.14$\times10^{20}$\,cm$^{-2}$ (Dickey \& Lockman \cite{Dickey}), and when the column
is frozen at this value, an equally good fit is obtained with no change in the other
parameters. This is indicative of the emission having no intrinsic absorption, \ie
lying, in some sense, above the host galaxy NGC~7793. This is good evidence for the
majority of this unresolved emission being truly diffuse gas.

\begin{figure}
\unitlength1.0cm 
\begin{picture}(8.5,12.0)
\label{fig_spec_diff}
\psfig{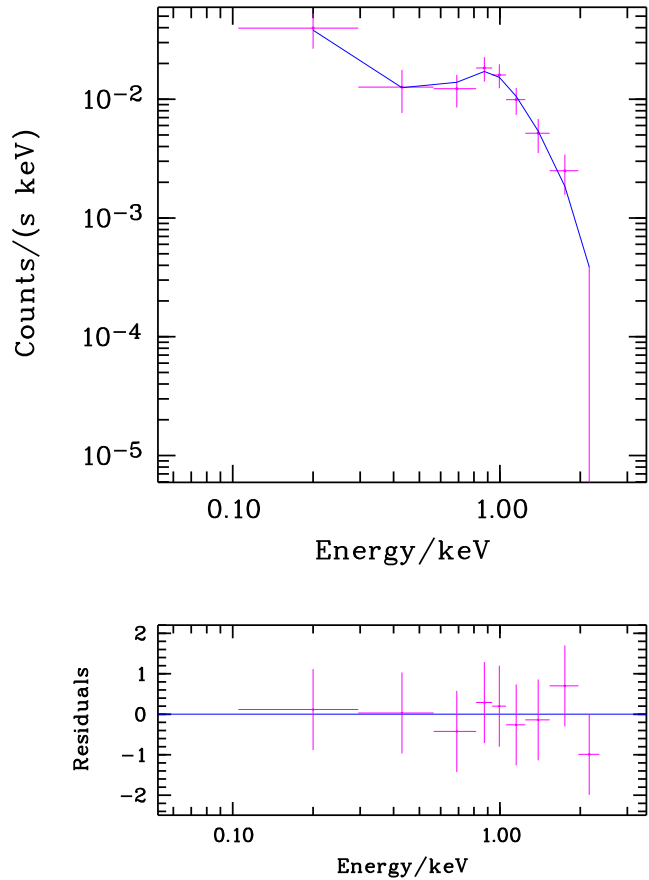}
\end{picture} 
\hfill \parbox[b]{8.7cm} 
{\caption{Unresolved emission spectrum with the best-fit Raymond \& Smith hot plasma
 plus absorption model
(see Table~\ref{table_fits_diff}). The pulse height spectrum of the total X-ray 
emission is indicated by crosses, and the model, by the solid line. 
}}
\end{figure}

\begin{figure}
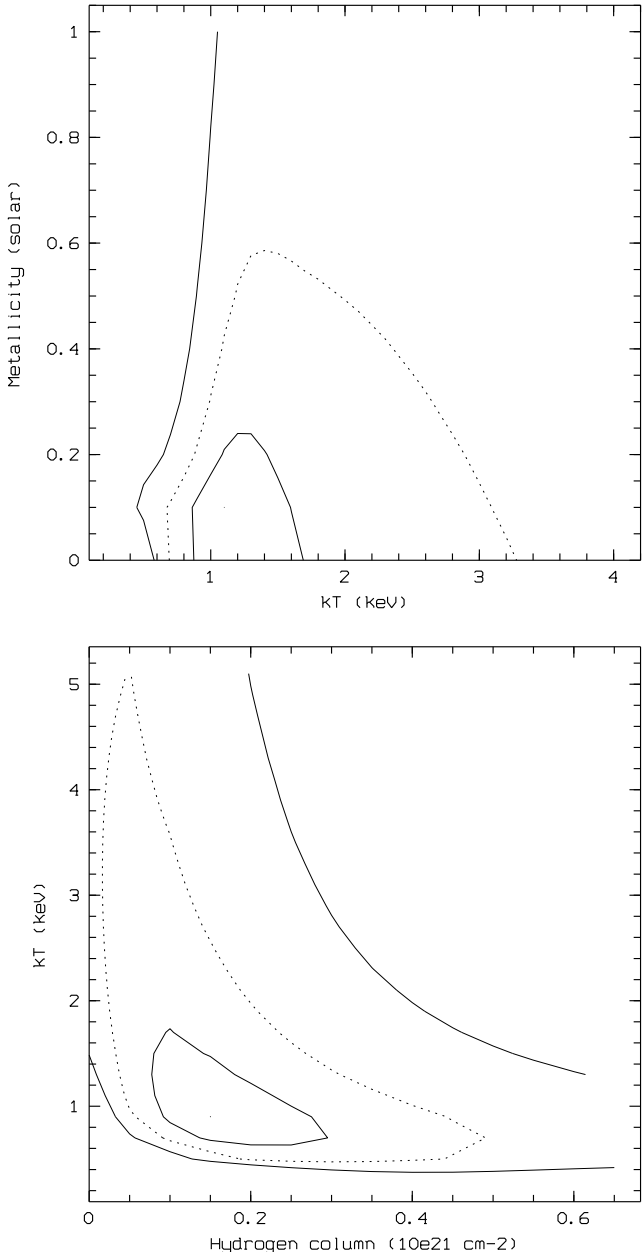

\unitlength1.0cm 
\begin{picture}(8.5,8.5)
\psfig{figure=7951.f9a,width=8.5cm,clip=}
\end{picture} 
\begin{picture}(8.5,8.5)
\psfig{figure=7951.f9b,width=8.5cm,clip=}
\end{picture} 
\hfill \parbox[b]{8.7cm} 
{\caption{Gaussian
contour levels of 1$\sigma$, 2$\sigma$ and 3$\sigma$ in the (top)
metallicity$-$temperature plane, and the (bottom) temperature$-$absorption
column plane for the Raymond \& Smith hot plasma model fit to the unresolved emission
spectrum.
}}
\end{figure}

The temperature of the unresolved emission appears to lie in the range 0.7$-$1.7\,keV,
with a best fit value of $\approx$1\,keV. This is rather on the high side when compared
with other galaxies, where the temperature of the diffuse emission has been estimated to
be more like 0.5\,keV or lower (\eg\ Bregman \& Pildis \cite{Bregman94}; Wang \etal\
\cite{Wang}; Ehle \etal\ \cite{Ehle}; Snowden \& Pietsch \cite{Snowden95}; Read \etal\
\cite{Read97PS}; Pietsch \& Vogler \cite{Pietsch98b}), and may indicate that some fraction of
the unresolved emission associated with NGC~7793 may be due to unresolved point sources.

It is here we must proceed with some caution. There are several threads of evidence
indicating that the unresolved emission is not truly diffuse, and is contaminated to
some degree with unresolved point source emission. Firstly, as discussed above, the
temperature is high. Secondly, as can be seen in Fig.\,5, even in the hard band, where
the unresolved emission contributes only a little, variations in the unresolved emission
structure can be seen between the December 1992 and the May 1993 observations (note
the changes in the `plumes' to the north-east and north-west of the bright southern P13
source). Whether these changes are significant is very difficult to say, as the 
features in question lie at the 2 or 3$\sigma$ level. Nevertheless, point-source 
contamination appears very possible. Comparison of equivalent soft band images to those
in Fig.\,5, where the unresolved emission (or at least the diffuse component of the
unresolved emission) contributes most, is fraught with difficulties, on account of
the large increase in the entire field-of-view background (mentioned earlier in
Sect.~\ref{sec_obs_data}) during the second half of the December 1992 observation.
Nothing very conclusive can be said regarding the variation in the unresolved
emission via this method.

In a further effort to establish whether there is any variation in the
unresolved emission, a similar procedure to that performed on the point
sources was performed here. Both the unresolved (the sources having been again
removed to radii of 0\farcm75 (1\farcm0 for P13)) and the background emission
were binned (using the same spatial regions as used in the extraction of the
equivalent spectra) into the two separate December 1992 and May 1993
observations, and also into the 14 individual observation blocks. The
unresolved emission was then corrected for background, vignetting effects and
exposure time, leading to unresolved emission count rates of 3.23$\pm$1.07 cts
s$^{-1}$ arcmin$^{-2}$ (December 1992) and 3.77$\pm$0.54 cts s$^{-1}$
arcmin$^{-2}$ (May 1993) (note the larger error in the December observation
due to the increased background contamination). This gives rise to a
probability of variability of 55.1\% (significant at only the 0.76$\sigma$
level). Furthermore, fitting of the 14-block lightcurve to a constant count
rate results in a good fit (with a $\chi^{2}$ of 14.15), suggesting that the
unresolved emission is only variable at a significance level of 0.91$\sigma$
($L=$1.01). The unresolved emission therefore appears to be non-variable in
terms of flux level, though this of course, does not exclude the possibilty of
the emission being contaminated with unresolved point sources.


Another way to search for contamination within the unresolved emission from
unresolved point sources, is to search for these sources spectrally. Several
two-component models were fitted to the unresolved emission spectrum. These models
had the general form of a cool ($\sim$0.5\,keV, more typical of spiral halos),
unabsorbed thermal component, representing the true diffuse gaseous emission, and a
hot ($\sim$10\,keV) component, representing any unresolved point source component.
Though all avenues were explored in terms of variation of models and freezing and
freeing of parameters, no significant improvement on the one-component RS model was
obtained.

Assuming at first then, that this unresolved emission is due entirely to hot gas,
then mean physical properties of this gas can be inferred from the above results if we
make some assumptions about the geometry of the emission. Here we have assumed the
simple geometry of the emission being hemispherical with a radius of 3\farcm5 (in
actuality, only a rough approximation to the gas properties can be calculated here
and assumption of a slightly different radius gives rise to very similar results).

Using the volume derived for this hemispherical `bubble' model, the fitted emission
measure $\eta n_{e}^{2} V$ (where $\eta$ is the `filling factor' -- the fraction of
the total volume $V$ which is occupied by the emitting gas) can be used to infer the
mean electron density, $n_{e}$, and hence the total mass $M_{\mbox{\small gas}}$,
thermal energy $E_{\mbox{\small th}}$ and cooling time $t_{\mbox{\small cool}}$ of
the gas.

Performing these calculations, after first accounting for the extra emission lost in
the `holes' left after the source-subtraction procedure, one arrives at approximate
values to the physical properties of the gaseous emission as follows; X-ray
luminosity $L_{X}$ (0.1$-$2.4\,keV) 4.1$\times10^{38}$\,erg s$^{-1}$; $n_{e}$,
4.6$\times10^{-3} \eta^{-0.5}$\,cm$^{-3}$; $M_{\mbox{\small gas}}$, $9.6\times10^{6}
\eta^{0.5}$\, $M_{\odot}$; $E_{\mbox{\small th}}$, 5.4$\times10^{55}
\eta^{0.5}$\,erg; $t_{\mbox{\small cool}}$, 4.2\,Gyr.
If, as seems likely however, some contamination from unresolved point sources exists, 
then the above results should only be taken as upper limits.

\section{NGC~7793 as a member of the Sculptor Group}
\label{sec_disc_n7793}

In comparing the X-ray properties of NGC~7793 with other galaxies, it is 
useful to compare its properties with those of its group neighbours.

The Sculptor group is possibly 
the nearest small group of galaxies to our own Local Group, and as such, much is
known about the individual members. Kinematical studies (\eg Puche \& Carignan
\cite{Puche88}) have established the group to be made up of five major members
(NGC~55, NGC~247, NGC~253, NGC~300 and NGC~7793). Other smaller galaxies, such as
NGC~24 and NGC~45, were found to be more distant, and although dwarf galaxies do
exist within Sculptor (\eg Lausten \etal\ \cite{Lausten}), they contribute
essentially nothing to the group emission or dynamics. Now, with the completion of
the present work, all five major Sculptor members have had their X-ray properties
analysed and the results presented (\eg\ Schlegel \etal\ \cite{Schlegel}; Read \etal\
\cite{Read97PS}; Vogler \& Pietsch \cite{Vogler}; Pietsch \& Vogler 
\cite{Pietsch98b}).
The \Ros\ PSPC observations of all four of NGC~7793's neighbours have been analysed
in a self-similar way, and are presented in Read \etal\ (\cite{Read97PS}).

Basic properties (including the present X-ray results) of the five Sculptor
galaxies are given in Table\,\ref{table_sculptor}, as follows; Col.\,1 gives
the galaxy name. Cols.\,2, 3 and 4 give the galaxy type, diameter and axis
ratio (all taken from de Vaucouleurs \etal's (\cite{RC3}) Third Reference
Catalog of Bright Galaxies). For each of NGC~7793's neighbours, the galaxy
distances quoted in Tully (\cite{Tully}) are given in col.\,5. These are based
on $H_{0} = 75$\,km~s$^{-1}$~Mpc$^{-1}$, and assume that the Galaxy is retarded
by 300\,km~s$^{-1}$ from universal expansion by the mass of the Virgo cluster.
Optical (B) luminosities (col.\,6) are taken from Tully (\cite{Tully}) and FIR
luminosities (col.\,7) are calculated from IRAS 60 and 100\,$\mu$m fluxes using
the expression

\begin{displaymath}
L_{FIR} = 3.65\times10^{5}\left[2.58 S_{60\mu m} + 
S_{100\mu m} \right] D^{2}  L_{\odot},
\end{displaymath}

(see \eg Read \& Ponman \cite{Read95P}). Here $D$ is the distance in Mpc and
$S_{60\mu m}$ and $S_{100\mu m}$, the IRAS 60 and 100\,$\mu$m fluxes (in Janskys),
are taken from Soifer \etal\ (\cite{Soifer}) and from Rice \etal\ (\cite{Rice}). The
(0.1$-$2.4\,keV) X-ray luminosities (from Read \etal\ (\cite{Read97PS}) and the
present paper) are given in col.\,8. We have here assumed, with reference to the
discussion in Sect.~\ref{sec_disc_sources}, that sources P6 and P7 are not
associated with NGC~7793. Two sets of values are given for NGC~7793 however, based
on the fact that the true nature of source P13 (the brightest in the field) is not
known. It may or may not be associated with NGC~7793.
Ratios of the 
three luminosities are given in cols.\,9, 10 and 11, and the temperature of 
the diffuse X-ray emission is given in col.\,12. The $T_{\rm diffuse}$ 
values for NGC~247, NGC~253 and NGC~300 are taken from Read \etal\
(\cite{Read97PS}), the value for NGC~55 is taken from Schlegel \etal\
(\cite{Schlegel}).

\begin{table*}
\caption[]{Properties of the Sculptor group galaxies (see text for details).
All X-ray information for NGC~7793 is taken from the present paper. All X-ray information
for the remaining systems is taken from Read \etal\ (\cite{Read97PS}), with the exception 
of the NGC~55 diffuse emission temperature, which is taken from Schlegel \etal\
(\cite{Schlegel}).}
\label{table_sculptor}
\begin{tabular}{lrrrrrrrrrrr}
\hline
\noalign{\smallskip}
Galaxy & Type & Diam. & Axis & Dist. & \multicolumn{3}{c}{Log luminosity (erg s$^{-1}$)} & 
 \multicolumn{3}{c}{Luminosity ratios ($10^{-4}$)} & $T_{\rm diffuse}$ \\
       &      & (kpc) & ratio& (Mpc) & $L_{B}$ & $L_{FIR}$ & $L_{X}$ & $L_{FIR}/L_{B}$ & $L_{X}/L_{B}$ 
 & $L_{X}/L_{FIR}$  & (keV) \\
(1) & (2) & (3) & (4) & (5) & (6) & (7) & (8) & (9) & (10) & (11) & (12) \\
\noalign{\smallskip}
\hline
\noalign{\smallskip}
NGC~55  & SBS9 & 12.3 & 5.8 & 1.3 & 43.02 & 41.94 & 38.91 & 830 & 0.78 & 9.3 & $\sim$1 \\
NGC~247 & SXS7 & 13.1 & 3.1 & 2.1 & 42.96 & 41.48 & 38.68 & 330 & 0.52 &16.0 & 0.16 \\
NGC~253 & SXS5 & 24.0 & 4.1 & 3.0 & 43.78 & 43.74 & 40.04 &9100 & 1.80 & 2.0 & 0.39 \\
NGC~300 & SAS7 &  7.6 & 1.4 & 1.2 & 42.52 & 41.43 & 38.18 & 810 & 0.46 & 5.6 & ($\sim$0.1) \\ \hline

NGC~7793&\raisebox{-1.4ex}{SAS7}&\raisebox{-1.4ex}{9.1} &\raisebox{-1.4ex}{1.5} 
        &\raisebox{-1.4ex}{3.4}& \raisebox{-1.4ex}{43.01} &\raisebox{-1.4ex}{42.23} 
        & 39.17 &\raisebox{-1.4ex}{1700}& 1.45 & 8.7 &\raisebox{-1.4ex}{\ltsim0.97} \\
(w/out P13) & & & & & & & 38.78 &                
            & 0.59 & 3.6 & 
             \\
\hline
\end{tabular}
\end{table*}

As one can see from Table~\ref{table_sculptor}, NGC~7793 is, like all the members of the
Sculptor group, a late-type spiral galaxy. It appears, in spite of its small size, to be the
second most active galaxy in the group, at least in terms of its far-infrared properties,
the most active being of course the famous large starburst, NGC~253. Disregarding NGC~253
for the moment, NGC~7793, although small, is as optically bright or brighter than the
remaining three, and has the highest $L_{FIR}$ value of the four non-starbursts.

In terms of NGC~7793's X-ray emission, the situation is complicated by the fact that P13
may or may not belong to the galaxy. If it does, then NGC~7793 appears to lie midway
between the very quiescent NGC~55, NGC~247 and NGC~300, and the starburst NGC~253, lying
directly on the $L_{X}:L_{FIR}$ correlation of Read \& Ponman (\cite{Read98}), and
slightly above their $L_{X}:L_{B}$ correlation. If P13 does not belong, then NGC~7793 
lies under the $L_{X}:L_{FIR}$ correlation, appearing
almost identical to M33 (Read \etal\ \cite{Read97PS}), and directly on the $L_{X}:L_{B}$
correlation, almost coincident with NGC~55. 

Why NGC~7793 is relatively rather bright (in all wavebands, possibly including the X-ray)
for its size may be due to that fact; its size, or rather, the size of its dark halo. In
an effort to explain why, when the average value of $(M/L_{B})_{\rm global} = 9 \pm 5
M_{\odot}/L_{\odot}$, for the individual Sculptor galaxies (the value for NGC~7793 itself
being the smallest, at around 5), the value for the group as a whole is ten times higher,
Puche \& Carignan, in a series of papers (\cite{Puche91} and the references therein),
produced \hi\ rotation curves and mass models for each of the five individual systems.
NGC~7793 is seen to possess the only truly declining rotation curve, and as such, is very
rare within the local universe (Carignan \& Puche \cite{Carignan90}). In fact, a model
comprising the luminous components and `no dark matter' can be adequately fit to the
NGC~7793 \hi\ data. Its halo is found to be very small ($r_{c} \approx 2.7$\,kpc,
approximately the size of the diffuse X-ray emission region), and its central density is
found to be a factor 10$-$15 higher than is seen in other late type spirals.

An interesting aspect of the X-ray emission from NGC~7793 is the unresolved emission, and
more specifically, its temperature. As mentioned earlier, the temperature obtained in the
spectral fitting of the unresolved spectrum, $\approx$1\,keV, is high, typical normal and
starburst galaxies containing diffuse emission at temperatures more like 0.5\,keV or
lower. Note (from Table~\ref{table_sculptor}) that the estimated temperature of the
`diffuse' emission in NGC~55, is also around 1\,keV (Schlegel \etal\ \cite{Schlegel}). As
we shall see here, comparing these two systems' unresolved emission properties is
particularly useful.

It is very possible, as discussed earlier, that the unresolved emission is due in part to
unresolved sources. We have discussed earlier our unsuccessful search for this unresolved
component, but perhaps it is worth some further discussion on more theoretical grounds.
Following similar arguments to those of Primini \etal\ (\cite{Primini}), and to those of
Schlegel \etal\ (\cite{Schlegel}) (who analysed the NGC~55 unresolved emission), one can
arrive at some interesting conclusions as regards the integrated emission of the
unresolved point sources (\ie\ stars, supernova remnants, cataclysmic variables and low-
and high-mass X-ray binaries) within NGC~7793.

Because of their very similar unresolved X-ray emission properties, both in terms of
luminosity (3.0$\times10^{38}$\,erg s$^{-1}$ for NGC~55, compared to 4.1$\times10^{38}$\,erg
s$^{-1}$ in the present case) and temperature, many of the conclusions reached as regards
NGC~55 by Schlegel \etal\ (\cite{Schlegel}), can be applied here to NGC~7793. NGC~7793's
smaller size however, means that the constraints placed are even more severe. For instance,
both cataclysmic variables (with typical $L_{X}$'s of $\sim10^{30-32}$\,erg s$^{-1}$;
Cordova (\cite{Cordova})) and active stars ($L_{X}\sim10^{31}$\,erg s$^{-1}$) fail, by
well over two orders of magnitude, to account for the unresolved X-ray emission within
NGC~7793. Supernova remnants ($L_{X}$, typically being $\sim10^{36}$\,erg s$^{-1}$) could
account for the emission, though a few hundred would be needed (compared to the 28
detected by Blair \& Long (\cite{Blair}). A further problem related to SNRs accounting for
the unresolved emission, is that, as NGC~7793 is rather face-on, the SNR spectra would be
relatively unabsorbed, and therefore too soft, when compared with the rather hard
unresolved spectrum observed.

X-ray binaries however, may well be able to explain the unresolved emission within 
NGC~7793. Only a hundred or so low-luminosity ($L_{X}\sim 10^{36-37}$\,erg s$^{-1}$)
systems, or just a handful of high luminosity ($L_{X}\sim 10^{38}$\,erg s$^{-1}$) 
systems are required to account for the flux seen.
Furthermore, the unresolved emission appears to be rather uniform (see Fig.\,2), and
follows the disc quite well, as would a population of evolved stars. 

One must be careful however as, for instance, if there were a largish
collection of XRBs within NGC~7793, we might expect to detect some, especially within
the disc. All of the discrete sources detected, with the exception of source P10 (which
is likely a SNR or superbubble), appear around the edge of the galaxy. No plausible XRB
candidates are detected within the NGC~7793 disc. Source P13 may possibly be a
high-mass XRB, though its luminosity is several times the Eddington limit for a
$1M_\odot$ compact object ($1.3\times 10^{38}$\,erg~s$^{-1}$). Furthermore, though the
temperature obtained for the unresolved emission is somewhat hotter than is generally
seen in spiral galaxies, it is not very hot, and is constrained, at least at the
1$\sigma$ confidence level to be less than $\sim1.7$\,keV. This may be too cool for the
emission to be explained in terms of a large proportion of it being due to evolved
stars (note the $2\sigma$ confidence limit to the unresolved emission temperature is
rather much high, however).

\section{Summary}
\label{sec_summary}

We have observed the \Ros\ PSPC data from a field centred on the nearby Sculptor
galaxy NGC~7793. 27 sources are detected within the central 25\arcm$\times$25\arcm
area, several appearing to be variable, and the brightest being situated to the
south of NGC~7793. In addition to point source emission, unresolved residual
emission is detected within and around NGC~7793. Our findings with regard to the
observed point-source and unresolved emission can be summarized as follows:

1. Seven point sources are detected within the optical confines of NGC~7793, two of
which appear to be unrelated to the host galaxy. Four of the others appear to be
associated with X-ray binaries, supernova remnants or superbubbles within the
galaxy. Interestingly, no significant enhancement is seen at the galactic nucleus.


2. The remaining source, the brightest point source in the field, lies on the
southern rim of NGC~7793, and is now seen not to be associated with a $z=0.071$
redshift QSO, as was previously thought. It may be associated with NGC~7793 (with a
(0.1$-$2.4\,keV) $L_{X}$ of 8.7$\times10^{38}$\,erg s$^{-1}$), though no obvious
counterpart is seen. It is seen to be significantly time variable, and may be a
high-mass, possibly black hole X-ray binary.

3. Unresolved residual emission is also observed, quite uniformly distributed over
the galaxy, extending to a radius of perhaps 4\,kpc. Spectral fitting of this
emission indicates that it is intrinsically, rather unabsorbed, suggesting that a
large component of the emission may be diffuse hot gas surrounding the system. The
fitted temperature however ($\approx1$\,keV) implies that a significant contribution
from unresolved point sources (here thought to be X-ray binaries) is likely.

4. Comparison of NGC~7793's X-ray properties with its Sculptor group neighbours (and
with other galaxies in general) depends largely on whether the bright source P13 is
associated with NGC~7793 or not. If it is, then NGC~7793, in terms of its activity,
lies somewhere between the very quiescent Sculptor galaxies NGC~55, NGC~247 and
NGC~300, and the well-known starburst NGC~253. If P13 is unrelated to NGC~7793, then
NGC~7793 appears rather similar to its quiescent neighbours.

\begin{acknowledgements}

AMR acknowledges the grateful receipt of a Royal Society fellowship during
this work. We are also very grateful to Thomas Boller for carefully 
reading this manuscript, to Andreas Vogler for his X-ray-optical
overlay routines, and to the referee, whose comments and suggestions have 
improved the paper. This research has made use of the SIMBAD database
operated at CDS, Strasbourg.

Optical images  are based on photographic data obtained with the UK Schmidt
Telescope, operated by the Royal Observatory Edinburgh, and funded by the UK
Science and Engineering Research Council, until June 1988, and thereafter by
the Anglo-Australian Observatory.  Original plate material is copyright (c)
the Royal Observatory Edinburgh and the Anglo-Australian Observatory. The
plates were processed into the present compressed digital form with their
permission. The Digitized Sky Survey was produced at the Space Telescope
Science Institute under US Government grant NAG W-2166.

Finally, we thank our colleagues from the MPE ROSAT group for their support.
The ROSAT project is supported by the German Bundesministerium f\"ur
Bildung, Wissenschaft, Forschung und Technologie (BMBF/DLR) and the
Max-Planck-Gesellschaft (MPG).

\end{acknowledgements}

\end{document}